\documentclass[twocolumn]{aastex631}

\newcommand{\lya}{Ly$\alpha$}

\newcommand{\change}[1]{#1}
\newcommand{\changerev}[1]{#1}

\usepackage{amsmath,amssymb,amsfonts}
\usepackage[caption=false]{subfig}
\usepackage[capitalise]{cleveref}
\usepackage{layouts}

\begin{document}

\title{An extremely metal-poor Lyman $\alpha$ emitter candidate at $z=6$ revealed through absorption spectroscopy}

\author[0000-0001-8986-5235]{Dominika {\v D}urov{\v c}{\'i}kov{\'a}}
\affiliation{MIT Kavli Institute for Astrophysics and Space Research, 77 Massachusetts Avenue, Cambridge, 02139, Massachusetts, USA}
\affiliation{Department of Physics, Massachusetts Institute of Technology, 77 Massachusetts Avenue Cambridge, MA 02139}

\author[0000-0003-2895-6218]{Anna-Christina Eilers}
\affiliation{MIT Kavli Institute for Astrophysics and Space Research, 77 Massachusetts Avenue, Cambridge, 02139, Massachusetts, USA}
\affiliation{Department of Physics, Massachusetts Institute of Technology, 77 Massachusetts Avenue Cambridge, MA 02139}

\author[0000-0003-3769-9559]{Robert A.\ Simcoe}
\affiliation{MIT Kavli Institute for Astrophysics and Space Research, 77 Massachusetts Avenue, Cambridge, 02139, Massachusetts, USA}
\affiliation{Department of Physics, Massachusetts Institute of Technology, 77 Massachusetts Avenue Cambridge, MA 02139}

\author[0000-0003-3174-7054]{Louise Welsh}
\affiliation{Centre for Extragalactic Astronomy, Department of Physics, Durham University, South Road, Durham DH1 3LE, UK}
\affiliation{INAF - Osservatorio Astronomico di Trieste, via G. B. Tiepolo 11, I-34143 Trieste, Italy}

\author[0000-0001-5492-4522]{Romain A.\ Meyer}
\affiliation{Department of Astronomy, University of Geneva, Chemin Pegasi 51, 1290 Versoix, Switzerland}

\author[0000-0003-2871-127X]{Jorryt Matthee}
\affiliation{Institute of Science and Technology Austria (ISTA), Am Campus 1, A-3400 Klosterneuburg, Austria}

\author[0000-0002-5360-8103]{Emma V.\ Ryan-Weber}
\affiliation{Centre for Astrophysics and Supercomputing, Swinburne University of Technology, Hawthorn, Victoria 3122, Australia}
\affiliation{ARC Centre of Excellence for All Sky Astrophysics in 3 Dimensions (ASTRO 3D), Australia}

\author[0000-0002-5367-8021]{Minghao Yue}
\affiliation{MIT Kavli Institute for Astrophysics and Space Research, 77 Massachusetts Ave., Cambridge, MA 02139, USA}

\author[0000-0003-1561-3814]{Harley Katz}
\affiliation{Department of Astronomy \& Astrophysics, University of Chicago, 5640 S Ellis Avenue, Chicago, IL 60637, USA}
\affiliation{Kavli Institute for Cosmological Physics, University of Chicago, Chicago IL 60637, USA}

\author[0000-0001-5818-6838]{Sindhu Satyavolu}
\affiliation{Institut de F{\'i}sica d’Altes Energies (IFAE), The Barcelona Institute of Science and Technology, Edifici Cn, Campus UAB, 08193, Bellaterra (Barcelona), Spain}

\author[0000-0003-2344-263X]{George Becker}
\affiliation{Department of Physics \& Astronomy, University of California, Riverside, CA 92521, USA}

\author[0000-0003-0821-3644]{Frederick B.\ Davies}
\affiliation{Max Planck Institut f\"ur Astronomie, K\"onigstuhl 17, D-69117, Heidelberg, Germany}

\author[0000-0002-6822-2254]{Emanuele Paolo Farina}
\affiliation{International Gemini Observatory/NSF NOIRLab, 670 N A’ohoku Place, Hilo, Hawai'i 96720, USA}

% \author[0000-0002-7054-4332]{Joseph F.\ Hennawi}
% \affiliation{Leiden Observatory, Leiden University, P.O. Box 9513, 2300 RA Leiden, The Netherlands}
% \affiliation{Department of Physics, University of California, Santa Barbara, CA 93106, USA}

\correspondingauthor{Dominika {\v D}urov{\v c}{\'i}kov{\'a}}
\email{dominika@mit.edu}



\begin{abstract}

We report the discovery of a Lyman $\alpha$ emitter (LAE) candidate in the immediate foreground of the quasar PSO J158-14 at $z_{\rm QSO}=6.0685$ at a projected distance $\sim29\ {\rm pkpc}$ that is associated with an extremely metal-poor absorption system. This system was found in archival observations of the quasar field with the Very Large Telescope/Multi-Unit Spectroscopic Explorer (VLT/MUSE) and was previously missed in searches of absorption systems using quasar absorption line spectroscopy as it imparts no detectable metal absorption lines on the background quasar spectrum. The detected \lya\ emission line at a redshift of $z_{\rm LAE}=6.0323$ is well aligned with the outer edge of the quasar's proximity zone and can plausibly cause its observed damping wing if it is associated with a proximate sub-damped \lya\ absorption system with a column density of $\log {N_{\rm HI} / {\rm cm}^{-2}} \approx 19.7$. A $>10$ hour medium-resolution spectrum of the quasar observed with the Magellan/Folded-port InfraRed Echellette (FIRE) and VLT/X-Shooter spectrographs reveals a metallicity constraint of ${\rm [Z/H]} < -3$. Such low metallicity makes this system an extremely metal-poor galaxy candidate and provides an exciting site to study possible signatures of Population III stars. 
\end{abstract}



\section{Introduction} \label{sec:intro}

Finding Population III (Pop III) stars is one of the most sought-after endeavors in modern astrophysics due to their important role in the formation of the first structures in the early Universe \citep[e.g.][]{klessen_first_2023}. Pop III stars are the first generation of stars that collapsed from pristine, (nearly) metal-free gas produced by the Big Bang nucleosynthesis. They are thought to create the first heavy elements and their remnants possibly provide the initial seeds for the earliest supermassive black holes (SMBHs). Their direct detection remains elusive as the majority of Pop III stars are thought to have mostly formed early in the history of the Universe ($z\sim 20-30$) and their likely heavy masses \cite[e.g.][]{abel_formation_2002,bromm_first_2017} caused them to have very short lifetimes ($\sim$ a few Myr). Furthermore, they are thought to reside in minihalos \citep{hirano_primordial_2015,schauer_influence_2019} with luminosities that should be challenging to detect at such high redshifts even with the sensitivity of the James Webb Space Telescope (JWST) \citep{schauer_ultimately_2020}.

Despite the challenges in directly detecting Pop III stars, significant progress has been made in studying their signatures indirectly through observational probes and theoretical modeling. Examples include the study of extremely metal-poor (${\rm [Fe/H]} < -3$) stars in the Local Group \citep[Pop II, e.g.][]{beers_discovery_2005,frebel_near-field_2015} or the imprints of Pop III stellar feedback on the largest scales \citep{madau_cosmic_2014}. Recently, substantial progress has been made on developing new optical emission-line diagnostics that could enable the study of this elusive stellar population in the early Universe \citep[e.g.][]{nakajima_diagnostics_2022, katz_challenges_2023,vanni_chemical_2024, katz_21_2024}. In fact, a handful of metal-poor systems at high-redshift have already been identified through their emission-line properties \citep[e.g.][]{vanzella_extremely_2023,cameron_nebular_2024,maiolino_jades_2024,cullen_jwst_2025,fujimoto_glimpse_2025,naidu_black_2025}, with metallicities as low as $12 + \log({\rm O/H}) < 6.3$, or $[{\rm O/H}]<-2.4$ \citep{vanzella_extremely_2023}.

% \cite{vanzella_extremely_2023} found a lensed metal-poor system at $z\sim 6.6$ with an emission-line based metallicity of $12 + \log(O/H) < 6.3$ ($[{\rm O/H}]<-2.4$).
% \cite{cameron_nebular_2024} identified a $z=5.9$ galaxy with a low metallicity of $12 + \log(O/H) = 7.59$ ($[{\rm O/H}]=-1.1$).
% \cite{maiolino_jades_2024} argue that the HeII emission of GN-z11 could be a sign of photoionization by Pop III stars.
% \cite{cullen_jwst_2025} reported a metallicity of $12 + \log(O/H) \approx 6.9$ ($[{\rm O/H}]\approx -1.8$) in a galaxy at $z\approx8.3$.
% \cite{fujimoto_glimpse_2025} discovered a lensed Pop III galaxy candidate at $z=6.5$ with a gas-phase metallicity of $[{\rm Z/H}]<-2.3$.
% \cite{naidu_black_2025} reported on a Little Red Dot whose SMBH is enshrouded in a dense, low-metallicity gas of $[{\rm Z/H}]\approx-2$.
% \cite{willott_search_2025} report a gravitationally-lensed galaxy at $z=8.2$ with an emission-line based metallicity of $12 + \log(O/H) =6.85$ ($[{\rm O/H}]=-1.84$)

Theoretical models predict that pristine gas pockets and Pop III star formation can persist beyond $z\sim6$ \citep{tornatore_population_2007,johnson_extreme_2019}. One method to search for such isolated environments is via absorption systems along sightlines to distant quasars 
\citep[e.g.][]{becker_high-redshift_2011,becker_evolution_2019,dodorico_evolution_2022,dodorico_xqr-30_2023}. Quasar absorption line spectroscopy has proven extremely useful in studying chemical enrichment across cosmic time \citep[e.g.][refer to \citeauthor{fan_quasars_2023} \citeyear{fan_quasars_2023} for a review]{simcoe_characterizing_2002,schaye_metallicity_2003,ryan-weber_downturn_2009,cooke_most_2011,davies_xqr-30_2023,welsh_survey_2024}, particularly through damped Lyman-$\alpha$ absorption systems (DLAs, $\log {N_{\rm HI} / {\rm cm}^{-2}} > 20.3$). DLAs produce distinctive features in a quasar spectrum. The neutral hydrogen produces a saturated Voigt absorption line profile with a Lorentzian damping wing in the \lya\ forest. This is accompanied by low-ionization metal absorption lines at the same redshift due to the metals present in the gas. More diffuse systems that are still optically thick to \lya, such as Lyman limit systems (LLSs) and sub-damped Lyman-$\alpha$ absorption systems (sub-DLAs) with $17.2 \leq \log {N_{\rm HI} / {\rm cm}^{-2}} \leq 20.3$, are of particular importance in searching for pristine, isolated environments as their densities are not sufficient to sustain star formation that would cause chemical enrichment by further generations of stars \citep{fumagalli_detection_2011,salvadori_first_2012,fumagalli_physical_2016,saccardi_evidence_2023}.

At $z\gtrsim5$, the increasingly neutral intergalactic medium (IGM) suppresses the \lya\ forest and renders identifying absorption systems therein significantly more challenging \citep[e.g.][]{gunn_density_1965,simcoe_interstellar_2020,fan_quasars_2023}. 
Searches for such high-redshift HI-rich gas clouds are thus mostly restricted to cases where the absorber lies in the region of increased flux transmission in the immediate foreground of the quasar, known as the proximity zone (e.g. \citeauthor{cen_quasar_2000} \citeyear{cen_quasar_2000}, \citeauthor{madau_earliest_2000} \citeyear{madau_earliest_2000}, \citeauthor{haiman_probing_2001} \citeyear{haiman_probing_2001}, \citeauthor{fan_constraining_2006} \citeyear{fan_constraining_2006}; also note recent work doing this in galaxy sightlines by \citeauthor{heintz_jwst-primal_2025} \citeyear{heintz_jwst-primal_2025}). Such systems have been found near a handful of high-redshift quasars as proximate DLAs \citep[pDLAs;][]{dodorico_witnessing_2018,banados_metal-poor_2019,davies_possible_2023}. Even then, disentangling the quasar's \lya\ damping wing due to the proximate absorber from the imprint of the neutral IGM has only been enabled through the detection of metal absorption lines in the quasar spectrum, which concurrently provide a valuable constraint on the metallicity of the intervening gas cloud \citep{banados_metal-poor_2019,simcoe_extremely_2012,wang_significantly_2020}. 

Given the required presence of metal absorption lines, the identified pDLAs near Reionization-era quasars are not pristine anymore, as they already show significant metal enrichment \citep{dodorico_witnessing_2018,banados_metal-poor_2019}. Therefore, to identify the most metal-poor systems that could potentially host Pop III stars, one has to look for proximate absorption systems in high-redshift quasar sightlines that do not produce strong associated metal line absorption. This is especially challenging during the Epoch of Reionization, as metal-poor proximate absorption systems can be easily confused with \lya\ damping wings caused by the IGM. In this letter, we report on the discovery of such a proximate absorption system in the field of a $z=6$ quasar.

\begin{figure*}[t!]
    \centering
    \includegraphics[width=0.95\linewidth]{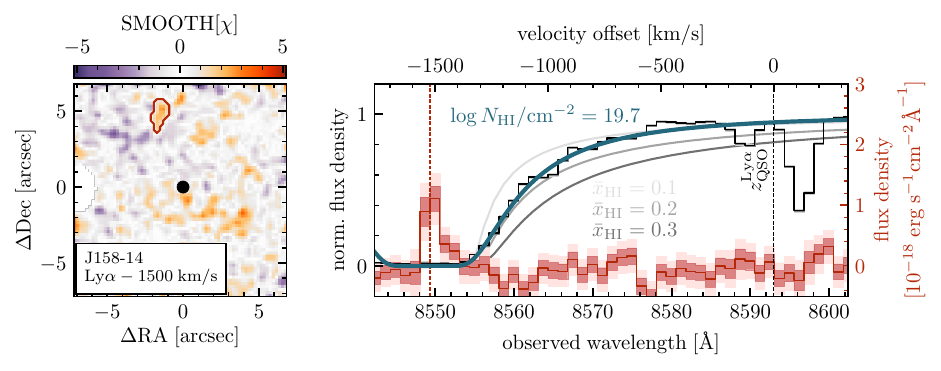}
    \caption{Left: A smoothed, PSF-subtracted SNR image (${\rm SMOOTH[\chi]}$) from MUSE showing the newly identified LAE in the foreground of the quasar PSO J158-14 (offset by $-1500\ {\rm km/s}$ from the \lya\ emission of the quasar). Right: The LAE spectrum from the original (not PSF-subtracted) MUSE data cube extracted over the contour shown in the left panel is shown in red ($1\sigma$ and $2\sigma$ uncertainties shown as shaded red regions). The quasar spectrum is shown in black with gray uncertainties, which are too small to be clearly visible, with its \lya\ redshift marked by $z_{\rm QSO}^{\rm Ly\alpha}$. The redshift of the LAE, $z_{\rm LAE}^{\rm Ly\alpha}=6.0323$ aligns with the end of the quasar's proximity zone, and an absorber with a column density of $\log {N_{\rm HI} / {\rm cm}^{-2}} \approx 19.7$ (a sub-DLA) can be used to model the damping wing of this quasar. For comparison, we also overplot example damping wing profiles arising from a homogenous IGM at a volume-averaged neutral gas fraction of $\bar{x}_{\rm HI} = 10\%, 20\%$ , and $30\%$ \citep{miralda-escude_reionization_1998}. Note that \change{the quasar spectrum has been continuum normalized to enable damping wing modeling (see \S~\ref{sec:metallicity} for details), while the LAE spectrum is shown in absolute flux density units (red y axis on the right)}. No other emission lines are seen across the wavelength range of MUSE.}
    \label{fig:LAE_MUSE}
\end{figure*}

\begin{figure}[h!]
    \centering
    \includegraphics[width=\linewidth]{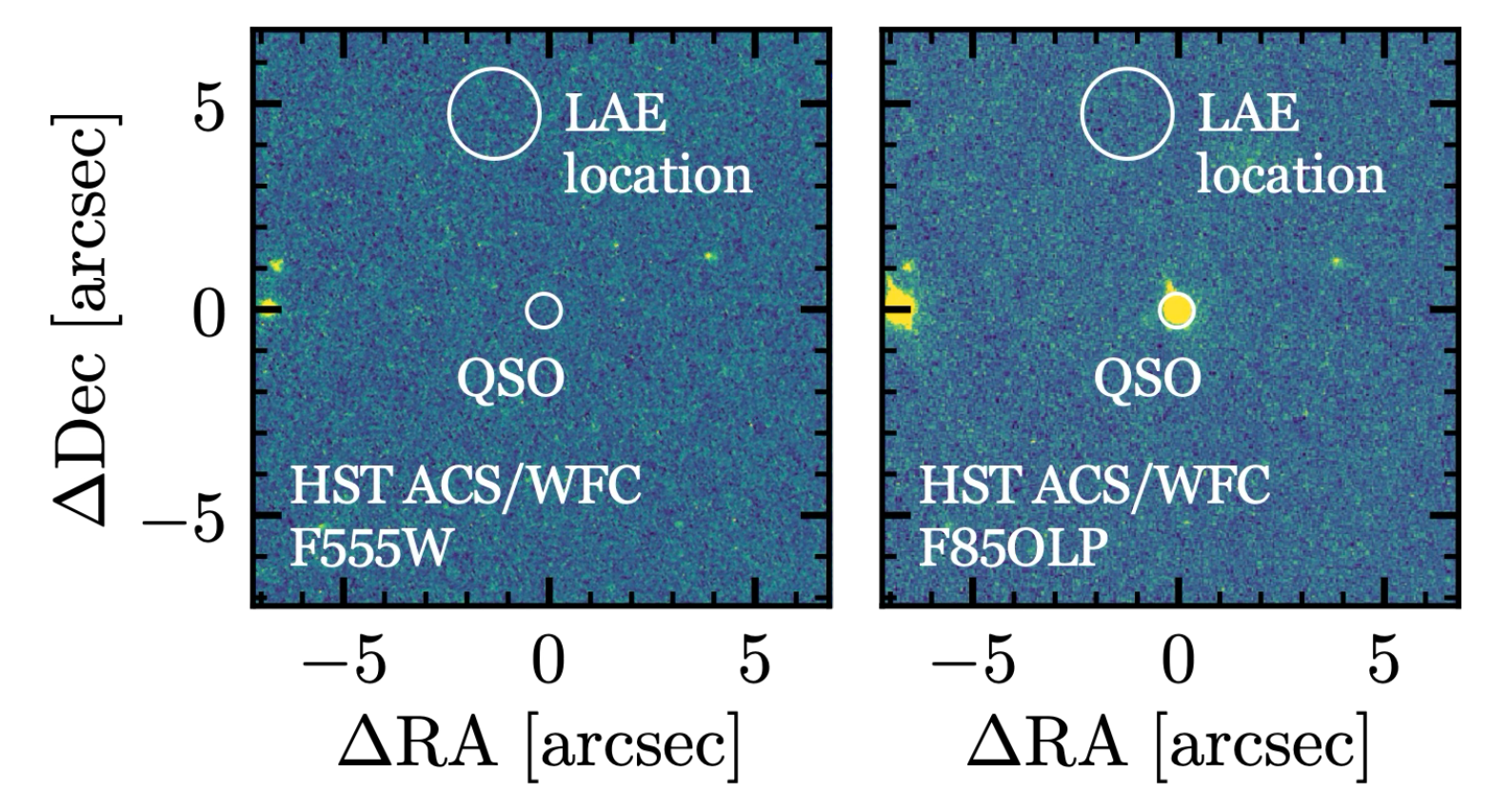}
    \caption{Archival imaging data of the quasar field from HST. Left: It is unlikely that this emission is not \lya\ as no foreground emission is detected by HST ACS/WFC imaging with the F555W filter at a limiting magnitude of $26.88$. Right: A non-detection in the F850LP filter at a limiting magnitude of $25.98$ is consistent with the \lya\ line flux measured in MUSE.}
    \label{fig:LAE_HST}
\end{figure}

Throughout this paper, we use the flat $\Lambda$CDM cosmology with $h = 0.67$, $\Omega_M=0.31$, $\Omega_\Lambda=0.69$ \citep{planck_collaboration_planck_2020}.

\section{Evidence for a metal-poor absorption system in PSO J158-14} \label{sec:data}

\begin{figure*}[t!]
    \centering
    \includegraphics[width=\linewidth,trim={0cm 0 -0.5cm 0.0cm},clip]{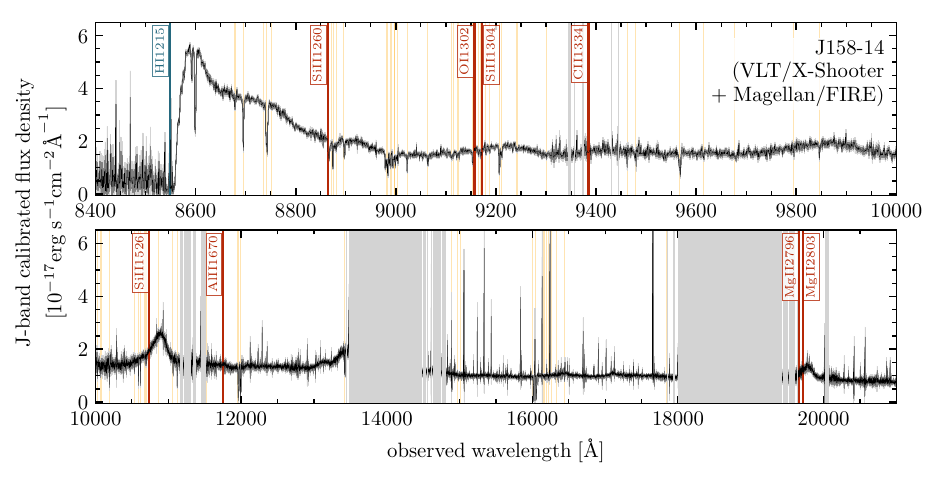}
    \includegraphics[width=\linewidth,trim={-0.5cm 0 0cm 0.5cm},clip]{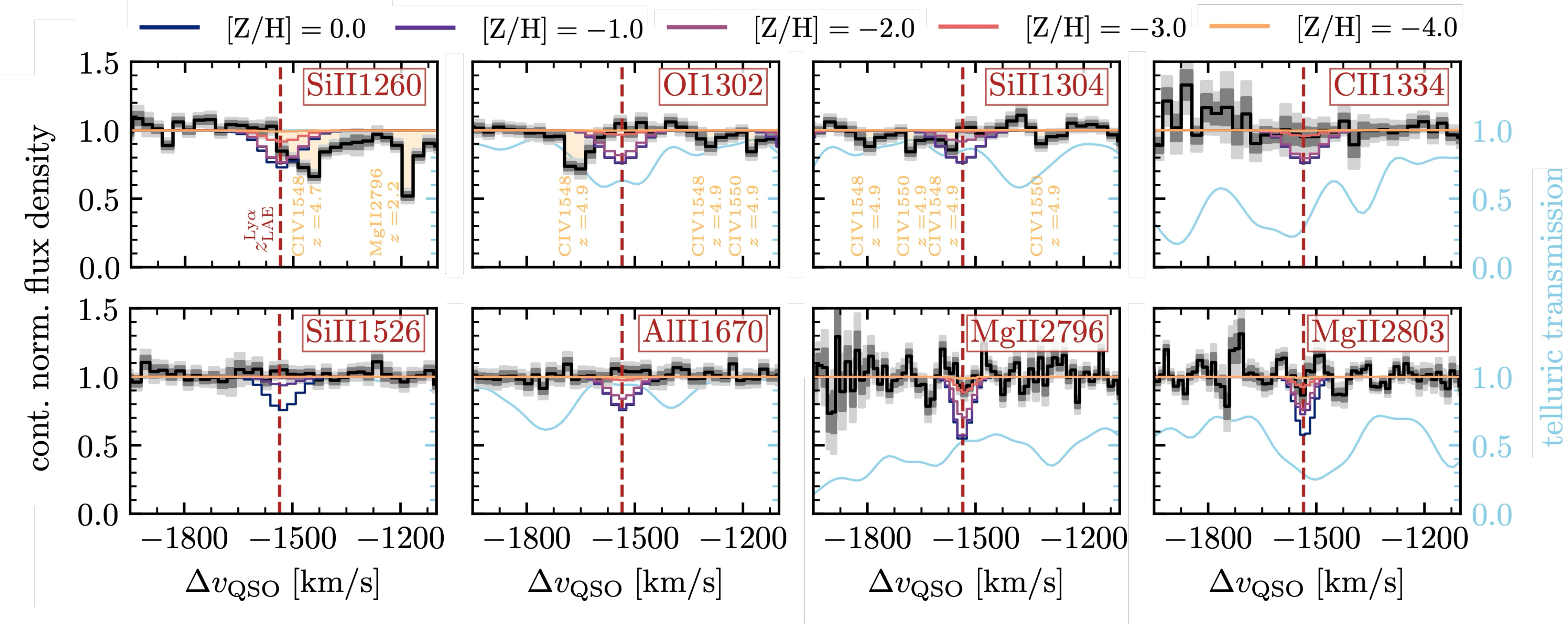}
    \caption{Top: A deep ($10.2\ {\rm hr}$) medium-resolution J-band calibrated spectrum of the quasar PSO J158-14, observed with the Magellan/FIRE and VLT/X-Shooter spectrographs. This spectrum was further continuum normalized (see main text) and used to stack regions corresponding to low-ionization absorption lines (shown as vertical red lines) that could be associated with the foreground LAE (marked by the blue vertical line). The orange vertical lines mark all absorption lines due to foreground systems identified by \cite{davies_xqr-30_2023}\change{, and the gray vertical lines mask spectral regions heavily contaminated by telluric emission.} Bottom: Zoom-in on the absorption line regions in the continuum-normalized spectrum that could be associated with the \change{proximate} absorber, in velocity relative to the quasar's rest frame. The dashed red line corresponds to the redshift of the LAE identified in MUSE observations, and orange regions mark absorption features due to other foreground systems \citep{davies_xqr-30_2023}. \change{Additionally, we show the telluric transmission in blue, and we overplot simulated absorption profiles at a range of metallicities $\rm [Z/H]$.} In both parts of the figure, the $1\sigma$ and $2\sigma$ uncertainties are shown as gray and light gray shaded regions.}
    \label{fig:spec}
\end{figure*}

\subsection{IGM damping wing or absorption system?}

The quasar PSO J158-14 at $z_{\rm QSO}=6.0685$ \citep{chehade_two_2018,eilers_detecting_2020} exhibits a somewhat unexpected damping-wing-like \lya\ transmission profile (\cref{fig:LAE_MUSE}). All current constraints on the history of reionization \citep{eilers_opacity_2018,bosman_new_2018,yang_measurements_2020,durovcikova_chronicling_2024,bosman_hydrogen_2022} imply a very low volume-averaged neutral gas fraction at this redshift ($\bar{x}_{\rm HI}\lesssim 0.2$), which makes seeing strong damping wings due to the IGM unlikely. There is growing evidence of neutral islands and damping wings persisting at $z<6$ \citep{becker_damping_2024,zhu_damping_2024,spina_damping_2024,sawyer_how_2025} due to a patchy reionization topology, however, simulations have shown that the likelihood of PSO J158-14 having a damping wing due to our sightline randomly crossing a neutral patch in the IGM is $<0.2\%$ \citep{satyavolu_new_2023}. \cite{satyavolu_new_2023} also noticed weak extended flux just blueward of the proximity zone indicating that this quasar's damping-wing-like absorption profile might in fact not be due to the IGM. Despite this fact, no metal absorption lines associated with a proximate absorption system have been found in the spectrum of this quasar \citep{eilers_detecting_2020,davies_xqr-30_2023}. This suggests that if a proximate absorber is the cause for the observed damping wing, it must be composed of metal-poor gas.

Using deep ($\sim8.5{\rm hr}$) 
observations with the Very Large Telescope/Multi-Unit Spectroscopic Explorer \citep[MUSE,][]{bacon_muse_2010}, we have identified a likely Ly$\alpha$ emitter (LAE) with an integrated line flux of $F_{\rm Ly\alpha}^{\rm LAE}\approx 2\times10^{-18}\ {\rm erg/s/cm^2}$ in the immediate foreground of the quasar PSO J158-14 at a transverse distance of $\sim5''$ ($\sim29\ {\rm pkpc}$\change{, with the coordinates of 10:34:46.6 $-$14:25:10.9}; left panel of \cref{fig:LAE_MUSE}). Note that the displayed pseudo-narrowband image (${\rm SMOOTH[\chi]}$\change{ is essentially a smoothed signal-to-noise image, following the definition of \citeauthor{farina_mapping_2017} \citeyear{farina_mapping_2017}}), collapsed over a spectral width of $100\ {\rm km/s}$, shows other patches of comparable SNR; however, only the pixels marked by the LAE contour show a clear, isolated emission line structure in the velocity space. At the position of this LAE, we detect no other emission lines over the wavelength range of MUSE. We refer the reader to {\v D}urov{\v c}{\'i}kov{\'a} et al. (2025\change{, in preparation}) for details on the MUSE observations and data reduction\change{ and here only note that the MUSE data were reduced using v2.8.7 of the MUSE Data Reduction Software \citep{weilbacher_design_2012,weilbacher_muse_2014} with standard parameters.}.

In order to test whether the detected emission line could be from a lower-redshift galaxy, we inspected archival imaging data of this quasar field taken with the Advanced Camera for Surveys (ACS) aboard the Hubble Space Telescope \citep[HST; Proposal ID: 16756, PI: Eilers; previously analyzed by][]{yue_detecting_2023}. Neither of the two $\sim30\ {\rm min}$ exposures, taken with the Wide Field Channel (WFC) F850LP and F555W filters, show a detection at the location of this LAE (\cref{fig:LAE_HST}). A non-detection in the F850LP image (which has a $3\sigma$ limiting AB magnitude of $25.98$ over the area of the LAE) is not surprising as the line flux detected from MUSE would correspond to a source with $m_{\rm AB} \approx 29.9$ in this image. \change{A non-detection of the LAE continuum further allows us to place a lower limit on the} observed equivalent width of \change{the \lya\ line of} ${\rm EW}_{\rm obs} ({\rm Ly\alpha}) \gtrsim 33\ {\rm \AA}$. A non-detection in the F555W image is also consistent with this emission line being \lya\ at $z\sim 6$ as its foreground flux would be suppressed by the IGM. If this source were, however, a foreground interloper, the low-redshift galaxy would have to be fainter than $26.88$ magnitude (which is the $3\sigma$ limiting AB magnitude over the LAE area)\footnote{Note that if the emission line detected in MUSE were H$\alpha$, this dwarf galaxy would be at $z=0.3$ and its [\ion{O}{2}]$3727{\rm \AA}$ emission line would fall into the F555W filter. Additionally, H$\beta$ and [\ion{O}{3}] lines would fall into the wavelength range of MUSE, and could have been detected. Additionally, if the detected emission line were [\ion{O}{2}]$3727{\rm \AA}$ (corresponding to a galaxy at $z\sim1.3$), we should be partially resolving its doublet shape. Therefore, also given the asymmetric profile of the observed line \citep{matthee_spectroscopic_2017,hayes_spectral_2021}, it is likely to be \lya\ at high-redshift.}.

Thus, we assume that the detected line is indeed \lya\ and the centroid of this emission line yields a redshift of $z_{\rm LAE}^{\rm Ly\alpha}=6.0323$, which aligns with the spectral region where we would expect an associated absorption system to be placed, i.e. the outer edge of the quasar's observed proximity zone (right panel of \cref{fig:LAE_MUSE}). Using a Voigt profile (shown in blue) \change{and assuming the absorber to be at the same redshift as the LAE}, we find that the observed absorption profile can indeed be modeled using a sub-DLA with an estimated \ion{H}{1} column density of $\log {N_{\rm HI} / {\rm cm}^{-2}} \approx 19.7$\changerev{, assuming the broadening parameter to be $b = 5\ {\rm km/s}$. This value of $b$ is in the range of typical values found in low-metallicity absorption systems \citep{cooke_most_2011,welsh_survey_2024,sodini_evidence_2024}, even considering that linewidths tend to be narrower as metallicity decreases \citep[e.g.][]{ledoux_velocity-metallicity_2006,murphy_connection_2007}. Note that the inferred \ion{H}{1} column density and the subsequent results are not sensitive to the precise choice of $b$ as long as $b\lesssim 50\ {\rm km/s}$, and the final metallicity constraint is unchanged even if $b$ is as low as $3\ {\rm km/s}$.}

\subsection{Metallicity from absorption spectroscopy}\label{sec:metallicity}

To derive a metallicity constraint for this absorber, we obtain a high-SNR, medium-resolution spectrum of the quasar to search for associated metal absorption lines. We co-added archival, shallow VLT/X-Shooter \citep{vernet_x-shooter_2011} observations of this quasar from January 24, 2017 ($4320\ {\rm s}/1.2\ {\rm hr}$; Program ID: 096.A-0418, PI: Shanks) with a newly observed deep spectrum taken with the Folded-port InfraRed Echellette (FIRE) spectrograph on the Magellan telescope \citep{simcoe_fire_2013}. Magellan/FIRE observations come from January 13, 2023 (9600 s), July 14, 2024 (4800 s), and Jan 13, 2025 (18000 s), totaling 9 hours of exposure time. These observations were reduced using the \texttt{PypeIt} package\footnote{\url{https://pypeit.readthedocs.io/en/latest/}} \citep{pypeit:joss_pub, pypeit:zenodo} and we refer the reader to \cite{durovcikova_chronicling_2024} for a more detailed description of the FIRE data reduction. The final spectrum, at a resolution of $\sim 30\ {\rm km/s}$, thus includes 10.2 hours of exposure time and is shown in the top part of \cref{fig:spec}. Spectra from individual nights were calibrated to $J_{AB}$ magnitude of 19.19 \citep{schindler_x-shooteralma_2020} before co-adding to create the final spectrum displayed here\footnote{The spectrum is made publicly available at \url{https://doi.org/10.5281/zenodo.15306220}}.

\begin{figure*}[t!]
    \centering
    \includegraphics[width=0.44\linewidth]{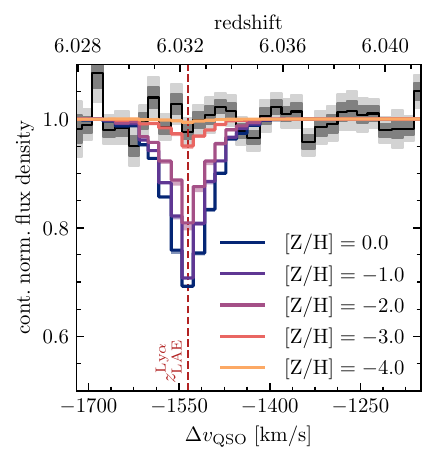}
    \includegraphics[width=0.55\linewidth,trim={0cm -0.6cm 0cm 0cm},clip]{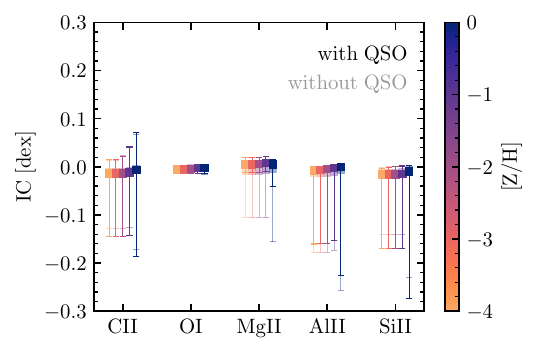}
    \caption{Left: Metallicity constraint on the newly found proximate absorption system. Here we show the \changerev{inverse-variance-weighted} stack of low-ionization absorption line regions in the continuum normalized spectrum of PSO J158-14, in terms of the velocity shift from the quasar's rest frame, shown in the bottom part of \cref{fig:spec} and also marked by vertical red lines in the top part of \cref{fig:spec}. The $1\sigma$ and $2\sigma$ uncertainties on the stacked spectrum are shown as gray and light gray shaded regions. \changerev{The modeled absorption profiles are stacked using the same weights as the data and overplotted as colored lines.} There is no detectable absorption at the redshift of the LAE at this spectral resolution, yielding a metallicity constraint of $[{\rm Z/H}] < -3$. This absorber thus constitutes an extremely metal-poor candidate. The shaded regions around the modeled absorption line profiles reflect the $1\sigma$ uncertainties on the ionization corrections derived from \textsc{CLOUDY}, as shown in the right panel. Right: Ionization corrections (with $1\sigma$ uncertainties from marginalizing over the modeled hydrogen densities) to the column densities of individual ions from \textsc{CLOUDY} used for absorption line profile modeling. Even though this absorption system is not fully self-shielding, the corrections are close to zero. This holds even when the radiation from the background quasar is taken into account (solid data points; compared to \textsc{CLOUDY} models without the quasar radiation shown as semi-transparent data points).}
    \label{fig:metallicity}
\end{figure*}

After fitting the quasar spectrum shown in the top part of \cref{fig:spec} and predicting its continuum emission around \lya\ following the methods in \cite{durovcikova_reionization_2020}, we normalize the observed spectrum by the predicted continuum to search for metal absorption lines. \change{We focus on regions of the continuum-normalized quasar spectrum where the accessible low-ionization metal lines at the redshift of the LAE would lie. We show these regions in the bottom part of \cref{fig:spec}, with the red dashed line marking the redshift where we would expect to find absorption lines associated with this LAE. To quantify the metallicity of this LAE, we convert our estimated \ion{H}{1} column density ($\log {N_{\rm HI} / {\rm cm}^{-2}} = 19.7$) to the column density of each atomic species assuming the solar abundance pattern \citep{asplund_chemical_2009}, and use the inferred column densities to forward model individual absorption line profiles for the aforementioned transitions using the assumed broadening of this absorption system ($b = 5\ {\rm km/s}$) at a range of metal abundance ratios $\rm [Z/H]$. The resultant profiles are overplotted in the bottom part of \cref{fig:spec}. We also note that some of the regions are contaminated by continuum fluctuations due to sky line contamination (the telluric transmission is shown in blue).}

\change{To derive a limit on the overall metallicity and increase the SNR, we proceed by stacking the absorption line regions and their associated profiles shown in the bottom panel of \cref{fig:spec}}. Note that during this procedure, we mask all absorption features in the spectrum that have previously been identified as corresponding to foreground intervening absorption systems \citep[shown as orange regions in \cref{fig:spec};][]{davies_xqr-30_2023}. \change{We show the inverse-variance-weighted mean stack} in black in the left panel of \cref{fig:metallicity}, with gray and light gray $1\sigma$ and $2\sigma$ uncertainties, respectively. \changerev{The modeled absorption profiles are stacked with the same weights as the data.} Note that the stacked spectrum is shown in terms of the velocity offset from the quasar -- therefore, any absorption associated with the LAE would be seen at the location of the red dashed line marking $z_{\rm LAE}^{\rm Ly\alpha}$. We detect no metal absorption lines at the redshift of the LAE beyond the noise level of the stacked, high-SNR spectrum, in agreement with previous studies that used shallower data \citep{eilers_detecting_2020, davies_xqr-30_2023}. 

As this system is a sub-DLA and therefore not fully self-shielding, we take into account ionization corrections \citep{viegas_abundances_1995} when converting the observed hydrogen column density to the column densities of the metals considered here,
\begin{equation}
        \log {N_{{\rm Z}_i}} = [{\rm Z/H}]+\log {N_{\rm HI}}+\log\left( \frac{N_{\rm Z}}{N_{\rm H}} \right)_\odot - {\rm IC}({\rm Z}_i).
\end{equation}
Here, $\log {N_{{\rm Z}_i}}$ is the column density corresponding to a particular ionization state of the metal ${\rm Z}$, $\log {N_{\rm HI}}$ is the column density of the neutral hydrogen (estimated from the quasar spectrum), $\log ( N_{\rm Z}/N_{\rm HI} )_\odot$ is the solar abundance ratio (from \citeauthor{asplund_chemical_2009} \citeyear{asplund_chemical_2009}), and ${\rm IC}({\rm Z}_i)$ is the ionization correction for a given ionization state at a given metallicity.

To obtain these ionization corrections, we ran a grid of \textsc{CLOUDY} \citep[v23.01,][]{ferland_cloudy_1998,chatzikos_2023_2023} models at a range of metallicities, $-4.0 \leq {\rm [Z/H]} \leq 0.0$ in steps of $1.0$, and hydrogen densities, $-3.0 \leq \log n_{\rm H}({\rm cm^{-3}}) \leq 3.0$ in steps of $0.5$. We include the UV background from \cite{haardt_radiative_2012} and the cosmic microwave background (CMB) at the redshift of the LAE, as well as cosmic rays. Additionally, we include the radiation of the background quasar as this could play a significant role in the case of this sub-DLA. \change{The quasar's spectral energy density is modeled via the \texttt{table AGN} command normalized to a bolometric luminosity of $L_{\rm bol} = 10^{47.31}\ {\rm erg/s}$ \citep{eilers_detecting_2020} and placed at a radius corresponding to the luminosity distance between the LAE and the quasar's systemic redshifts ($d_L\sim 15.9\ {\rm pMpc} $).} We do not include the effects of dust grains as these are known to be negligible at the low metallicities most relevant to this system \citep{de_cia_metals_2018,vladilo_evolution_2018,hamanowicz_metal-z_2024}. From these models, we derive ${\rm IC}({\rm Z}_i)$ by comparing the total abundance of the element $Z$ with respect to hydrogen to the ratio of the neutral hydrogen column density to the column density of the ionization state under consideration \citep{viegas_abundances_1995}. Subsequently, we marginalize over the grid of modeled hydrogen densities to obtain a distribution of ionization correction for each ionic species across the different metallicities. 

The derived ionization corrections are small ($\lesssim 0.3\ {\rm dex}$), especially at lower metallicities, even when the irradiation by the background quasar is taken into account (right panel of \cref{fig:metallicity}; results from \textsc{CLOUDY} runs without the quasar radiation are shown for comparison as semi-transparent data points). This is comparable to other sub-DLA corrections found in the literature \citep{berg_sub-damped_2021,welsh_survey_2024}. Taking these corrections (including their $1\sigma$ uncertainties) into account yields an upper limit on the metallicity of this absorber of ${\rm [Z/H]} < -3$ (left panel of \cref{fig:metallicity}). Such metallicity constraint implies an extremely metal-poor absorption system in the foreground of PSO J158-14, consistent with the redshift of the foreground LAE at a projected distance of $\sim29\ {\rm pkpc}$. 

% \cite{vanzella_extremely_2023} found a lensed metal-poor system at $z\sim 6.6$ with an emission-line based metallicity of $12 + \log(O/H) < 6.3$ ($[{\rm O/H}]<-2.4$).
% \cite{cameron_nebular_2024} identified a $z=5.9$ galaxy with a low metallicity of $12 + \log(O/H) = 7.59$ ($[{\rm O/H}]=-1.1$).
% \cite{maiolino_jades_2024} argue that the HeII emission of GN-z11 could be a sign of photoionization by Pop III stars.
% \cite{cullen_jwst_2025} reported a metallicity of $12 + \log(O/H) \approx 6.9$ ($[{\rm O/H}]\approx -1.8$) in a galaxy at $z\approx8.3$.
% \cite{fujimoto_glimpse_2025} discovered a lensed Pop III galaxy candidate at $z=6.5$ with a gas-phase metallicity of $[{\rm Z/H}]<-2.3$.
% \cite{naidu_black_2025} reported on a Little Red Dot whose SMBH is enshrouded in a dense, low-metallicity gas of $[{\rm Z/H}]\approx-2$.
% \cite{willott_search_2025} report a gravitationally-lensed galaxy at $z=8.2$ with an emission-line based metallicity of $12 + \log(O/H) =6.85$ ($[{\rm O/H}]=-1.84$)

\section{Conclusion}\label{sec:conclusion}

Using integral field unit spectroscopy, we identified a new LAE in the immediate foreground of the quasar PSO J158-14 at $z\approx 6$ that likely represents an extremely metal-poor proximate absorption system at ${\rm [Z/H]} < -3$. We detect no associated metal absorption lines in a deep, $10.2$-hour medium-resolution spectrum of the quasar. 

This high-redshift system is unique in that its low metallicity constraint is based on absorption spectroscopy, unlike other metal-poor galaxies identified at high-redshift with emission-line-based metallicities from JWST \citep{vanzella_extremely_2023,cameron_nebular_2024,maiolino_jades_2024,cullen_jwst_2025,fujimoto_glimpse_2025,naidu_black_2025,willott_search_2025}. \change{The metallicity limit derived here is} lower than the most metal-poor high-redshift galaxies identified to date with metallicity constraints of $[{\rm O/H}]<-2.4$ \citep{vanzella_extremely_2023} and $[{\rm Z/H}]<-2.3$ \citep{fujimoto_glimpse_2025}\change{. However, it should be noted that emission and absorption line spectroscopy typically each probe different gas phases, which makes it difficult to directly compare their inferred metallicities \citep[e.g.][]{james_investigating_2014,hernandez_first_2021}}.

The existence of this foreground LAE further implies that the proximity zone of the quasar is likely truncated, which in turn implies that the estimate of this quasar's lifetime based on the extent of its proximity zone is underestimated. This is supported by the fact that the MUSE data reveal a large \change{($24_{-9}^{+24}{\rm pkpc}$) \lya\ nebula} around PSO J158-14 ({\v D}urov{\v c}{\'i}kov{\'a} et al. 2025\change{, in preparation}).

In order to confirm the redshift and the low metallicity of this absorption system, deep, high-resolution spectroscopic observations of the quasar are required to resolve such weak metal lines (as demonstrated by \citeauthor{welsh_towards_2023} \citeyear{welsh_towards_2023} in lower-redshift metal-poor DLAs). Detecting these possible absorption features would also enable us to expand this analysis beyond the assumptions of the solar abundance pattern. Additionally, at a projected distance of $\sim29\ {\rm pkpc}$, the quasar sightline is likely probing gas outside of the virial radius of this LAE, suggesting that we might be probing its circumgalactic medium that has not been enriched. Therefore, in order to confirm the low-metallicity nature of its host galaxy candidate, direct spectroscopic observations of the LAE itself are necessary. If its metallicity is confirmed, this newly detected source could prove to be the most metal-poor system currently known at high redshift and thus provide a unique opportunity to search for signatures of Pop III stars. 

\section*{Acknowledgments}

\change{We thank the referee for their feedback and suggestions that greatly improved the quality of this manuscript.}

We would like to thank Carlos Contreras, Matías Díaz, Carla Fuentes, Mauricio Martínez, Alberto Pastén, Roger Leiton, Hugo Rivera, and Gabriel Prieto for their help and support during the Magellan/FIRE observations. We would also like to thank Rongmon Bordoloi for helpful discussions. RAM acknowledges support from the Swiss National Science Foundation (SNSF) through project grant 200020\_207349.

This paper includes data gathered with the 6.5 meter Magellan Telescopes located at Las Campanas Observatory, Chile. 

Based on observations collected at the European Southern Observatory under ESO programs 106.215A and 096.A-0418.

\change{The HST data presented in this article were obtained from the Mikulski Archive for Space Telescopes (MAST) at the Space Telescope Science Institute. The observations analyzed in this work can be accessed via \dataset[doi: 10.17909/gxmz-zd87]{https://doi.org/10.17909/gxmz-zd87}.}


\vspace{5mm}




% \appendix


\bibliography{LAE_refs}{}
\bibliographystyle{aasjournal}



\end{document}